\documentclass{revtex4-1}
\usepackage{amsmath}
\usepackage{amsfonts}
\usepackage{amssymb}
\usepackage[pdftex]{color,graphicx}
\usepackage{hyperref}
\usepackage{subfigure}
\usepackage{multirow}
\graphicspath{{figures/}}

\begin{document}

\title{High-Power Tunable Laser Driven THz Generation in Corrugated Plasma Waveguides}

\author{Chenlong Miao}
\affiliation{Institute for Research in Electronics and Applied Physics, University of Maryland, College Park, MD 20742, USA}
\author{John P. Palastro}
\affiliation{Naval Research Laboratory, Washington DC 20375, USA}
\author{Thomas M. Antonsen}
\affiliation{Institute for Research in Electronics and Applied Physics, University of Maryland, College Park, MD 20742, USA}
\date{\today}

\begin{abstract}
The excitation of THz radiation by the interaction of an ultra short laser pulse with the modes of a miniature corrugated plasma waveguide is considered. The axially corrugated waveguide supports the electromagnetic (EM) modes with appropriate polarization and  subluminal phase velocities that can be phase matched to the ponderomotive potential associated with laser pulse, making significant THz generation possible. This process is studied via full format Particle-in-Cell (PIC) simulations that, for the first time, model the nonlinear dynamics of the plasma and the self-consistent evolution of the laser pulse in the case where the laser pulse energy is entirely depleted. It is found that the generated THz is characterized by lateral emission from the channel, with a spectrum that may be narrow or broad depending on the laser intensity. A range of realistic laser pulse and plasma parameters is considered with the goal of maximizing the conversion efficiency of optical energy to THz radiation. As an example, a fixed drive pulse (0.55 J) with a spot size of 15 $\mu m$ and duration of 15 $fs$ produces 37.8 mJ of THz radiation in a 1.5 cm corrugated plasma waveguide with an on axis average density of $1.4\times10^{18} cm^{-3}$. 
\end{abstract}

\maketitle

\section{Introduction}\label{intro}
Terahertz radiation (THz) lies between microwave and infrared wavelengths in the electromagnetic spectrum and spans frequencies from 300 GHz to 20 THz. A wide variety of applications for THz radiation \cite{thzReport} can be found including time domain spectroscopy (TDS) \cite{Nuss1997}, medical and biological imaging \cite{mittleman1999}, remote detection \cite{mittleman1996, nusinovich2011range, isaacs2016remote} among others. Most airports, for instance, use millimeter wave/THz scanners for security checking. Existing small scale THz sources based on laser-solid interaction are limited to $\mu J$/pulse levels due to material damage \cite{Budiarto1996} although a recent discovery using optical rectification (OR) \cite{Vicario:14, Wu:15, Fulop:14} in organic crystals can exceed this limit. The need for higher power sources has led to the consideration of THz generation via laser-plasma interactions  \cite{PhysRevLett.91.074802, kim2008coherent, PhysRevLett.111.074802}, in which THz peak energies of tens of $\mu J$ can be achieved. Higher energy THz pulses can be generated at accelerator facilities via synchrotron \cite{PhysRevLett.63.1245} or transition radiation  \cite{PhysRevLett.67.2962}. Such facilities are relatively large and expensive to operate. This motivates research interest in small-scale, high efficiency terahertz sources.

Research has been actively conducted to investigate THz radiation generation by laser pulses propagating in plasmas since it was first demonstrated by Hamster et al. \cite{PhysRevLett.71.2725, hamster1994short}. In this case, the source of the radiation is the current driven by the ponderomotive force of a laser pulse. However, generation of radiation by a laser pulse propagating through uniform plasma is generally minimal. In order for electromagnetic modes to efficiently couple to the driving source, which travels at the group velocity of the laser pulse, the plasma must be inhomogeneous or immersed in a strong background magnetic field. A THz generation scheme involving laser pulses (or possibly electron beams) propagating through axially corrugated plasma channels has been proposed by Antonsen et al. \cite{antonsen2007excitation, antonsen2010radiation}. In this scheme the corrugated plasma channel acts as a slow wave structure that supports electromagnetic modes that can be phase matched with the driver. The scheme offers the possibility of much higher efficiency of THz generation under conditions of full laser pulse depletion. This prospect is investigated here.

\begin{figure}
\begin{center}
\includegraphics[width=16.0cm]{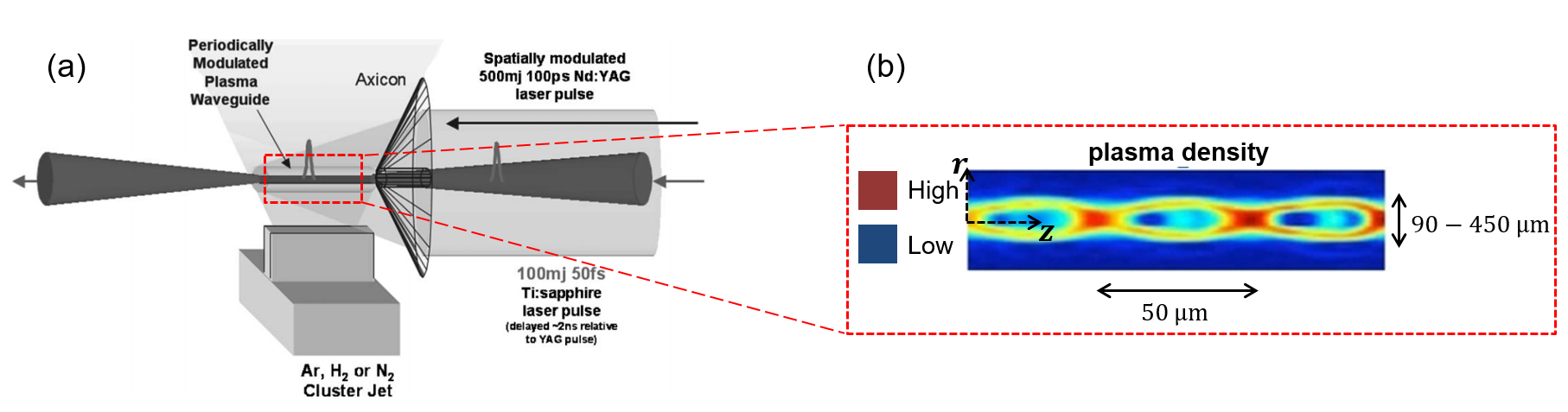}
\end{center}
\caption{(Color Online) (a) Diagram of an experimental setup for generating an axially corrugated plasma channel \cite{PhysRevLett.99.035001, HineSLM16}. (b) Snapshot of an experimentally generated modulated plasma channel (only 3 periods of many are shown).}\label{exp_setup}
\end{figure}

In this paper we report theoretical, numerical and simulation results of ponderomotively driven THz generation by laser pulses propagating through corrugated plasma waveguides. Such waveguides have already been realized in the laboratory \cite{PhysRevLett.99.035001, HineSLM16}. The experimental set up is shown in Fig.~\ref{exp_setup}a. A Nd:YAG laser pulse is line-focused onto a cluster jet creating a plasma that hydrodynamically expands, leading to the formation of a channel with a transversely parabolic density profile. The periodic structure is created by spatially modulating the laser intensity on the back side of the axicon, or by periodically obstructing the cluster flow. A second ultra-short Ti:Sapphire laser pulse is then injected into the channel following the channel formation pulse. The second pulse drives the terahertz generation. Shown in Fig.~\ref{exp_setup}b is a snap shot of the experimentally generated axially modulated plasma density profile. 

The organization of this paper is as follows. In Sec.~\ref{principles}, we introduce the mechanisms underlying THz generation in corrugated plasma waveguides. A mathematical model of the channel is considered and the dispersion relation is analyzed. In Section.~\ref{simulation}, we conduct full format PIC simulations to investigate the self-consistent evolution of the laser pulse in the strongly nonlinear regime and optimize the conversion efficiency from laser pulse to THz energy.  Tunability of the THz spectrum is also discussed. A numerical method to approximately find the frequencies of modes and coupling to the vacuum is also provided. The dependence of THz radiation on plasma density, driver pulse intensity, laser pulse duration, channel length and other channel parameters is investigated and discussed in detail. Two different types of plasma channels are considered and compared. In Section.~\ref{conclusion} we present our conclusions and discuss future directions.

\section{Excitation of THz modes in corrugated plasma channels}\label{principles}

\subsection{Ponderomotive Driver}
We consider the interaction of a drive pulse with a tenuous plasma. The central frequency of the drive pulse is significantly higher than the generated electromagnetic waves. The pulse produces a cycle-averaged low frequency ponderomotive force on the electrons, which induces an electron current that can produce radiation. The force on electrons is the gradient of the ponderomotive potential of the pulse, $\boldsymbol{F_p}=-\boldsymbol{\nabla}V_p$, where $V_p=mc^2a^2/4$, the normalized laser vector potential is $a=eE_a/mc\omega_0$, $m$ is the electron mass, $e$ is the electron charge and $c$ is the speed of light in vacuum, $\omega_0$ is the laser carrier frequency and $E_a$ is the electric field amplitude with both spatial and temporal dependence. The plasma ions are relatively heavy and can be treated as a stationary background during the interaction. In order for the excited current to generate electromagnetic radiation, the plasma must be inhomogeneous so that in no reference frame are the driven currents static.

Further to have significant THz emission, the phase velocity of the excited channel modes must match the group velocity of the laser pulse in the plasma. In a uniform plasma, the modes typically have superluminal phase velocities according to the dispersion relation $\omega^2=k^2c^2+\omega_p^2$, where $\omega_p=\sqrt{4\pi e^2N/m}$ is the plasma frequency. However, the phase velocity in a corrugated plasma channel can be subluminal, allowing phase matching. In particular, the periodic spatial modulations of the channel parameters cause the electromagnetic modes of the channel to have a Floquet type representation, which implies subluminal partial waves \cite{PhysRevE.83.056403}. Because of the superposition of spatial harmonics, these EM modes can have components with subluminal phase velocities. Thus, significant THz emission can be achieved.

\subsection{Corrugated Plasma Channels}\label{channel}
\begin{figure}
\begin{center}
\includegraphics[width=16.0cm]{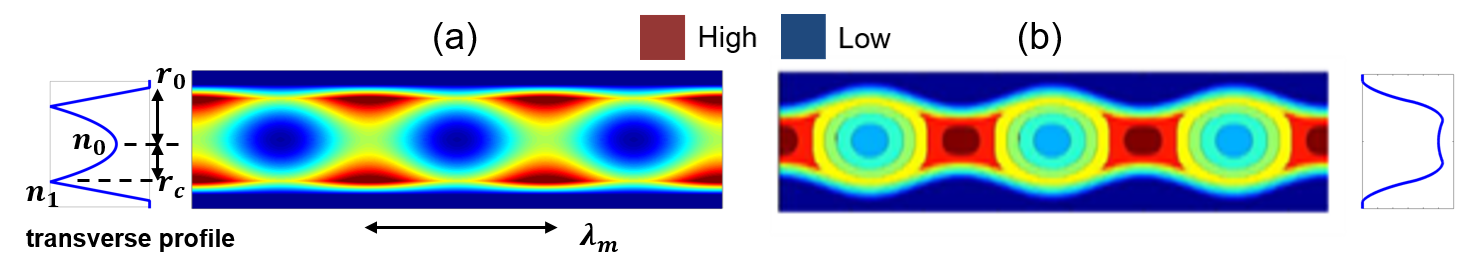}
\end{center}
\caption{(Color Online) False color image of two example electron density profiles generated by Eq.~(\ref{density_profile}): (a) an ``off-axis peak" channel with maximum density at the lateral edges. (b) an ``on-axis peak" channel with maximum density at the center. Here red is high density and blue is low density. All parameters of the two types of channels are displayed in Table~\ref{table_parameters}. In (b), both cut-off radius and channel radius are also modulated as $r_c[\mu m]=22.5-7.5\sin{(k_mz)}$ and $r_0[\mu m]=r_c+15$. Radially average density profiles of each type of channels are shown on the left and right, respectively.}\label{density_profile3}
\end{figure}

We consider the corrugated plasma waveguides to be cylindrically symmetric, with electron densities $N(r,z)$ described by the following,

\begin{equation}\label{density_profile}
  \frac{N(r,z)}{N_{0}} = \left\{
    \begin{array}{cc}
      \displaystyle{n_0+(n_1-n_0)\frac{r^2}{r_c^2}} \qquad & r \le r_c\\
     \displaystyle{ n_1 \frac{r_0-r}{r_0-r_c}} \qquad &r_c < r < r_{0}\\
      \displaystyle{\phantom{\frac{1}{1}}0} \qquad &r \ge r_{0}
    \end{array} \right.
\end{equation}
where $N_{0}$ is a normalization density. The quantities $n_0$, $n_1$, $r_0$ and $r_c$ are all potentially periodic functions of $z$ with period $\lambda_m$ and the modulation wavenumber $k_m=2\pi/\lambda_m$. The quantity $n_0(z)$ is the normalized on-axis density and $n_1(z)$ is the normalized density at $r=r_c$. For our studies we take $n_0=1+\delta \sin{(k_mz)}$ and $n_1 = \overline{n}_1+\delta_{1}\sin(k_mz)$, respectively, $\delta$ is the density modulation amplitude of the on-axis density $n_0$, $\overline{n}_1$ is the average density at $r=r_c$ and $\delta_1$ is the density modulation amplitude of $n_1$. The density has a parabolic transverse profile to guide the laser pulse during propagation. The quantity $r_c$ is the radius at which the density reaches $n_1$, the density then decreases linearly to zero from $r_c$ to $r_0$. The quantities $r_c$ and $r_0$ may also be axially modulated, $r_{c,0}=\overline{r}_{c,0}+\Delta_{c,0} \cos(k_mz+\theta_{c,0})$. Figures~\ref{density_profile3}a and \ref{density_profile3}b are false color images of two different density profiles, based on parameters given in the caption. Both profiles are similar to those realized experimentally \cite{PhysRevLett.99.035001, HineSLM16}. The channel in Fig.~\ref{density_profile3}a has its peak density off axis, while the channel in Fig.~\ref{density_profile3}b has its peak density on axis. We refer to these two types as the ``on-axis peak'' (Fig.~\ref{density_profile3}b) and the ``off-axis peak'' (Fig.~\ref{density_profile3}a) channels, respectively.

\begin{table}
\begin{center}
\caption{Typical parameters*.}\label{table_parameters}
\begin{tabular}{c|*{10}{c}}
\hline \hline
\multirow{3}{*}{laser pulse} &\multicolumn{3}{c}{central wavenumber} & \multicolumn{2}{c}{spot size} & \multicolumn{2}{c}{pulse duration} &\multicolumn{3}{c}{normalized vector potential}  \\
 &\multicolumn{3}{c}{$\lambda$}& \multicolumn{2}{c}{$r_L$} & \multicolumn{2}{c}{$\tau$, FWHM} & \multicolumn{3}{c}{$a_0$}   \\
 &\multicolumn{3}{c}{800 nm} &\multicolumn{2}{c}{15 $\mu m$} & \multicolumn{2}{c}{50 fs}  & \multicolumn{3}{c}{0.4} \\
\hline
channel type & $\lambda_m$[$\mu m$] & $\overline{r}_c$[$\mu m$] & $\overline{r}_0$[$\mu m$] & $\delta$ & $\overline{n}_1$ & $\delta_1$ & $\Delta_c$ & $\theta_c$ & $\Delta_0$ & $\theta_0$ \\
 off-axis peak (Fig.~\ref{density_profile3}a) & 15  & 30  & 40 & 0.9 & 3 & 0.9  & 0 & 0 & 0 & 0 \\
 on-axis peak (Fig.~\ref{density_profile3}b) & 15  & 22.5 & 37.5 & 0.7 & 1.3 & 0.1  & 7.5 & $\pi$/2  & 7.5 & $\pi$/2 \\
\hline \hline
\multicolumn{11}{c}{*Note: unless otherwise stated, all parameters are identical to those in Table~\ref{table_parameters}.}
\end{tabular}
\end{center}
\end{table}

\subsection{Dispersion Relation and Mode Excitation}\label{dispersion}
We review the linear theory of channel modes presented in Ref.~\cite{PhysRevE.83.056403}. The plasma is taken to be a cold fluid with linear response. Radially polarized, azimuthally symmetric TM modes $(E_r, E_z, B_\theta)$ of the channel are considered. In this case, an approximate wave equation assuming $\left | \boldsymbol{\nabla} \cdot \boldsymbol{E}\right |<<\left | \boldsymbol{E}\right |/r_c$ \cite{antonsen2007excitation,PhysRevE.83.056403} for the radial electric field $E_r$ can be derived
\begin{equation}\label{wave_equation}
	\bigg(-\frac{1}{c^2}\frac{\partial ^2}{\partial t^2}+\frac{\partial ^2}{\partial z^2}+\frac{1}{r}\frac{\partial}{\partial r}r\frac{\partial}{\partial r}-\frac{1}{r^2}\bigg) E_r=\frac{\omega _{p0}^2}{c^2}\frac{N(r,z)}{N_{0}}E_r\; ,
\end{equation}
where $\omega _{p0}$ is the plasma frequency evaluated for the normalization density $N_{0}$.

In the case in which the electron density profile is parabolic as $r\to \infty$, i.e. it is characterized by the first line of Eq.~(\ref{density_profile}), and we take $\delta_1=\delta$, we can find analytic expressions for the eigenmodes. Specifically, the $\gamma$th order radial eigenmode of the channel is described by
\begin{equation}\label{Er_eigen}
	E_r(r,z,t)=E_0f(z)H_\gamma \left( r/w_{ch}\right)\exp{\left(-\frac{r^2}{w_{ch}^2}\right)}\exp{(-i\omega t)}\; ,
\end{equation}
where $w_{ch}$ is the mode width given by $4/w_{ch}^4=\omega_{p0}^2(n_1-n_0)/{r_c^2c^2}$. The function $H_\gamma$ is the $\gamma$th polynomial defined as $\sum_{n=1}^{n=2\gamma-1} \alpha_n\left(r/w_{ch}\right)^n$ with coefficients determined by $\alpha_n/\alpha_{n-2}=4(n-1-2\gamma)/(n^2-1)$ and $\alpha_1=1$. Given the dependence of the fields on the radial coordinates, the function $f(z)$ satisfies the following Mathieu equation,
\begin{equation}\label{fz_eqn}
	\frac{d^2f}{dz^2}+k_0^2f=\frac{\omega _{p0}^2}{c^2}\delta \sin{(k_mz)}f \;,
\end{equation}
where the value of $k_0$ is related to the frequency by
\begin{equation}\label{channel_disp}
	\omega ^2-\left(\omega_{p0}^2 +\frac{8\gamma c^2}{w_{ch}^2}\right)= k_0^2c^2\; ,
\end{equation}
The dispersion relation is found by solving Eq.~(\ref{fz_eqn}) with Floquet boundary conditions, $f(z+\lambda_m)=\exp{(ik_z\lambda_m)}f(z)$. This then determines the dependence $k_0(k_z)$, which is inserted in Eq.~(\ref{channel_disp}) and gives $\omega(k_z)$ the dispersion relation.

\begin{figure}
\begin{center}
\includegraphics[width=8.0cm]{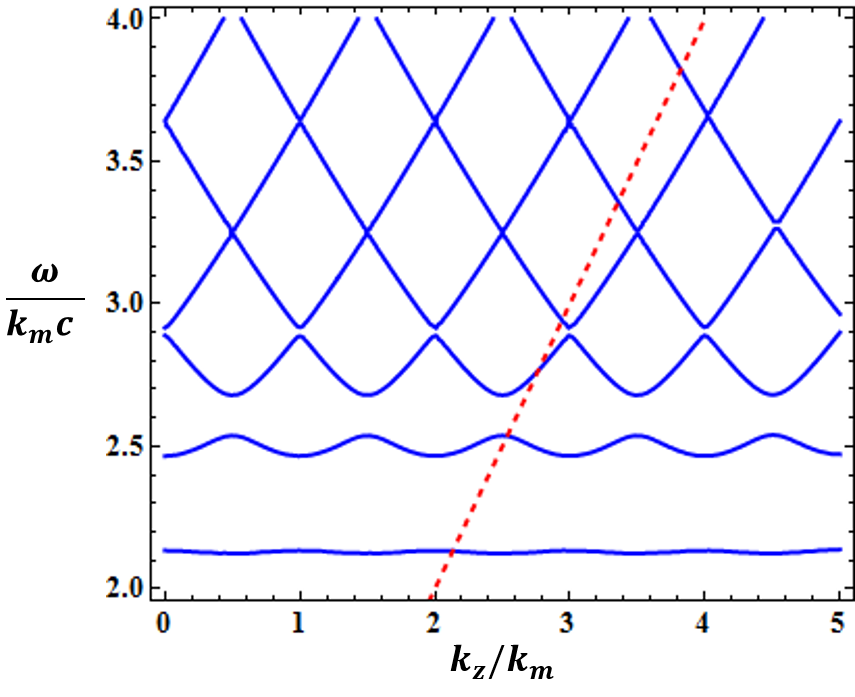}
\end{center}
\caption{(Color Online) Dispersion relation curves of the lowest (fundamental) radial mode of a corrugated channel evaluated by Eq.~(\ref{channel_disp}) with $N_{0}=1.4\times10^{18} ~cm^{-3}$, $\delta_1=\delta=0.9$, $\overline{n}_1=3$ and $r_c=30~\mu m$. The straight line (red) corresponds to the laser pulse moving at the speed of light, $\omega=k_zc$.}\label{dispersion_plot}
\end{figure}

Shown in Fig.~\ref{dispersion_plot} is the dispersion relation for the lowest ($\gamma=1$) radial mode of the model channel with parameters given in the caption. The dependence of $\omega$ on phase advance demonstrates the characteristic periodicity of frequencies in $k_z$ for periodic structures. The laser pulse, represented by a straight line in the plot, moves at its group velocity ($v_g\simeq c$) in the plasma channel. At places where the dashed pulse line and the dispersion relation curve intersect, phase matching occurs, and THz excitation can be expected at these frequencies.

Equations~(\ref{fz_eqn}) and (\ref{channel_disp}) apply as long as $w_{ch}<r_c$, that is, the THz or laser spot size is smaller than the channel width. Equation~(\ref{channel_disp}) is a good approximation to the dispersion relation regarding the cycle-averaged density profile and assuming the transverse parabolic shape extends to infinity. However, to determine the exact frequencies of the excited modes, one can numerically evaluate the wave equation, Eq.~(\ref{wave_equation}) using an exact electron density profile. A more accurate calculation of the dispersion relation, including the comparison with Eq.~(\ref{channel_disp}) is provided in Appendix \ref{appendix1}.

\section{Simulation Results}\label{simulation}

\subsection{THz Mode Excitation}\label{mode_excitation}
\begin{figure}
\begin{center}
\subfigure{
 \includegraphics[width=8.0cm]{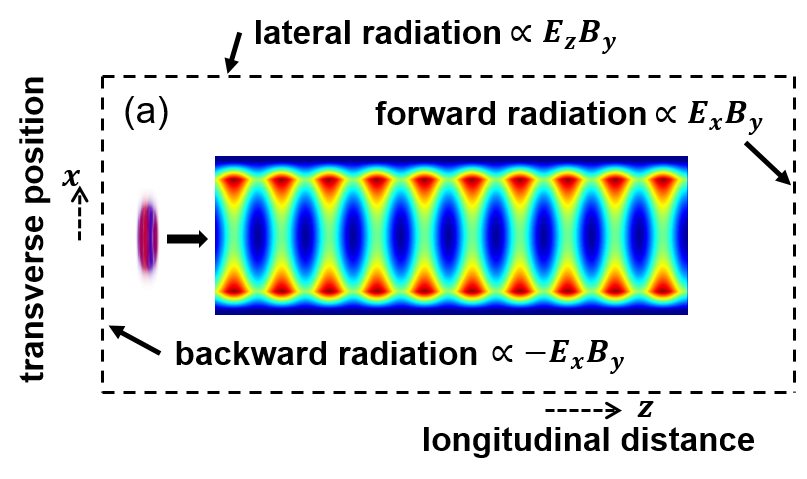}\label{simulation_setup}
}
\subfigure{
 \includegraphics[width=8.0cm]{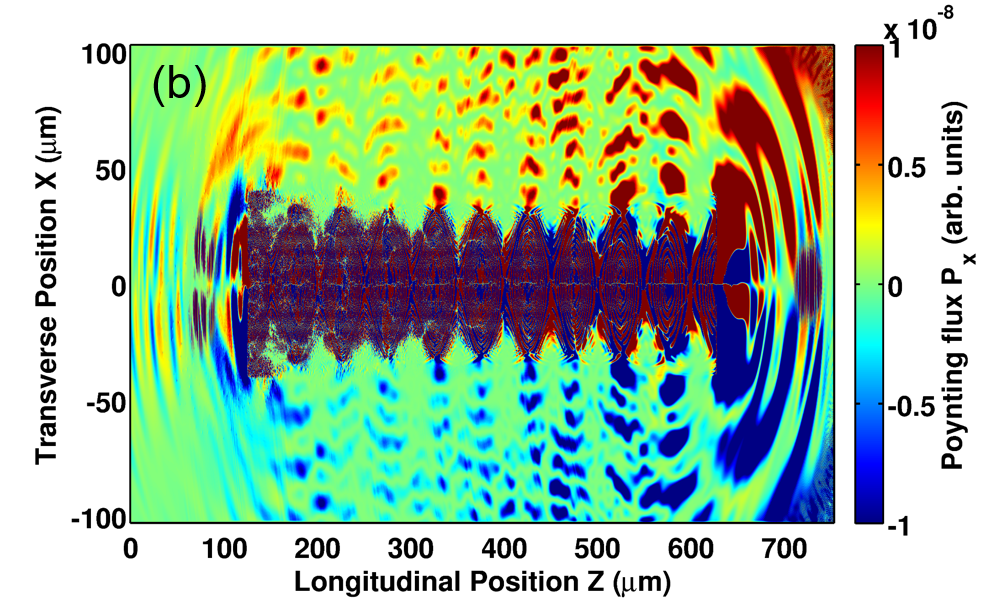}\label{THz_snapshot}
}  	
\end{center}
\caption{(Color online) \subref{simulation_setup} Diagram of simulation set up. The diagnostic box is set outside the plasma channel and the Poynting flux through each surface is calculated. The plasma channel consists of 10 modulation periods with $N_{0}=1.4\times10^{18} ~cm^{-3}$ and all other parameters in Table~\ref{table_parameters}.  \subref{THz_snapshot} A snap shot of the transverse component of Poynting flux $E_zB_y$ in PIC simulations after the laser propagates through the channel from left to right, with red and blue streaks indicating lateral THz radiation is generated.}
\end{figure}

THz generation in corrugated plasma waveguides is simulated using the full format PIC code TurboWAVE \cite{gordon2007}. The simulations, performed in 2D planar geometry, feature a finite sized plasma channel illuminated by an ultra short, intense laser pulse incident from the simulation boundary. Figure~\ref{simulation_setup} shows an example of an off-axis density peak plasma channel of 10 periods with modulation wavelength $\lambda_m=50~\mu m$. To quantify the radiation emitted from the plasma channel, we calculate the Poynting flux and its spectral density through each of the surfaces outside the plasma region indicated by the dashed lines. This captures emission in the forward, backward and lateral directions. The initial 2D simulations are performed in the lab frame in a domain of dimensions $205.9\times753.8$ $\mu m$ with $1024\times20480$ cells in the $x$ and $z$ directions, respectively. A laser pulse, with parameters detailed in Table \ref{table_parameters}, traverses the plasma channel from left to right. The initial normalized vector potential is $a_0=0.4$ (pulse energy $U_L=66~mJ$) with $a_0$ defined as $a_0=eE_0/mc \omega_0$, where $E_0$ is the peak field amplitude. Figure~\ref{THz_snapshot} shows a false color image of the transverse component of Poynting flux $P_x\propto E_yB_z$ after the laser pulse traverses the channel. The excited plasma wave can be observed as the rapid oscillations of $P_x$ inside the channel; the alternate positive and negative values indicate a small average flux. However, at both lateral boundaries (top and bottom of the image), one can observe the THz emission in the form of the red and blue streaks in the lateral Poynting flux.

\begin{figure}
\begin{center}
\includegraphics[width=10.0cm]{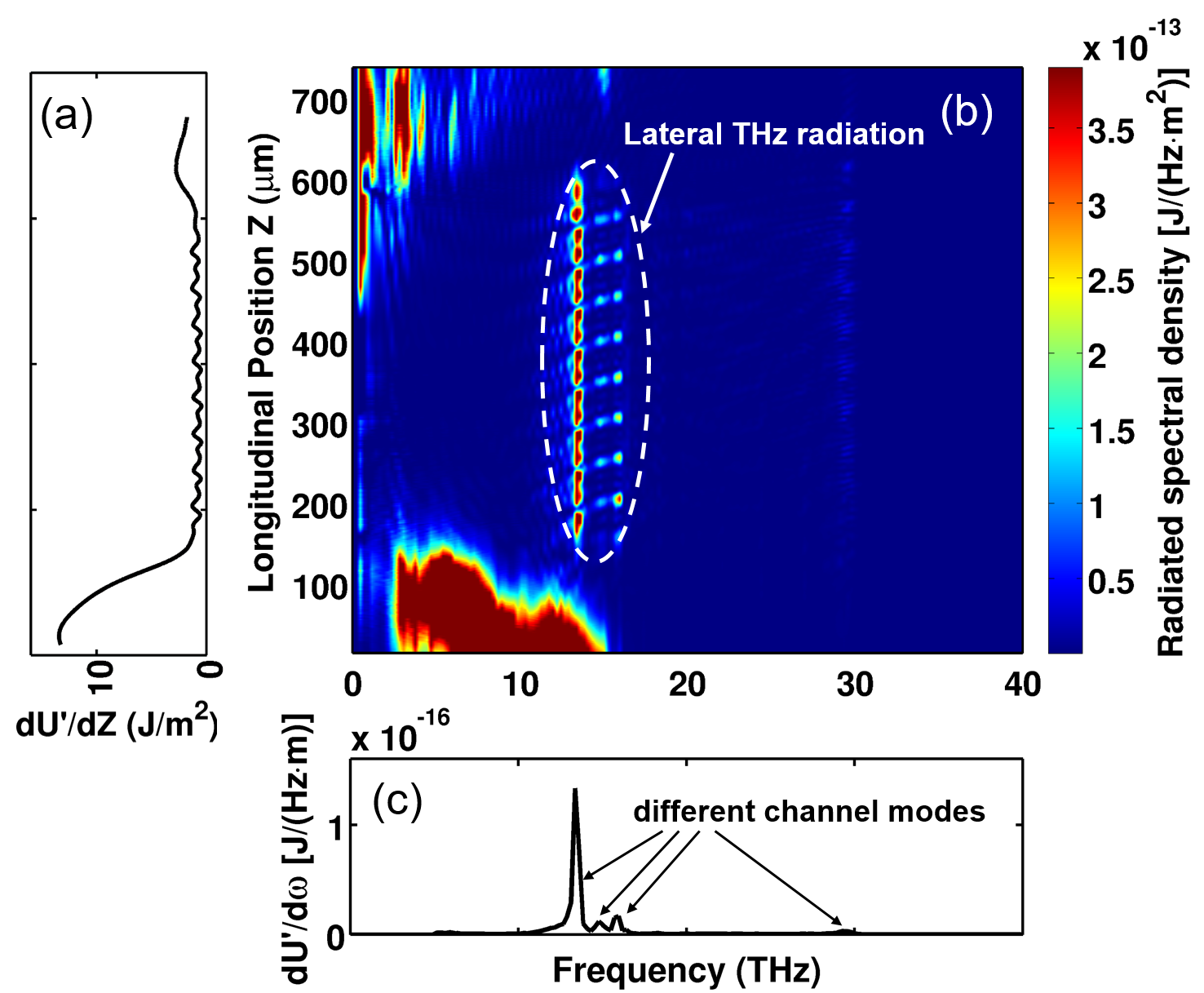}
\end{center}
\caption{(Color Online) (a) Radiated THz energy across the lateral diagnostic boundary shows two different mechanisms of generating THz. (b) Simulation results of radiated THz spectral density across the lateral diagnostic boundary using the same channel parameters in Fig.~\ref{simulation_setup} and $a_0=0.4$. Besides the low frequency, broad band THz radiation when laser pulse crosses the plasma interface \cite{chenlongRTR}, lateral THz radiation is also observed. (c) Radiated THz spectrum (only channel modes are considered in this plot) shows different channel modes are excited. The frequency of each excited mode matches well with the phase matching condition in Fig.~\ref{dispersion_plot}. }\label{SWPM_simu1} 
\end{figure}

To investigate the spectrum of the lateral THz emission, the fields are Fourier transformed in time. The radiated energy per unit length $U'$ through each diagnostic surface in Fig.~\ref{simulation_setup} can be obtained in the form of a spectral density in $(\omega,\boldsymbol{r})$, where $\boldsymbol{r}$ denotes the 2D spatial coordinates $(x,z)$:
\begin{subequations}
\begin{align}
\frac{dU'}{d\omega}&=\int dA \boldsymbol{\hat{n}} \cdot \boldsymbol{S}(\omega, \boldsymbol{r})\\
U'&=\int_{0}^{\infty} d\omega \frac{dU'}{d\omega}\;,
\end{align}
\end{subequations}
where $\boldsymbol{\hat{n}}$ is the unit vector normal to the surface $A$. The spectral density is given by
\begin{equation}
\label{spectral_density}
\boldsymbol{S}(\omega,\boldsymbol{r})=\frac{c}{8\pi^2}\big(\boldsymbol{E}(\omega,\boldsymbol{r}) \times \boldsymbol{B}^*(\omega,\boldsymbol{r})+c.c.\big)\; .
\end{equation}
To quantify the THz radiation emitted across each diagnostic surface, we calculate the $z$ component of the spectral density, $S_z$, for the left and right diagnostic boundaries and the $x$ component, $S_x$, for the lateral boundary. Figure~\ref{SWPM_simu1}(b) is the laterally radiated spectral density, $S_x$, from a PIC simulation of the off-axis type density profile. The low frequency, broad band THz radiation observed at the entrance of the channel, shown in Fig.~\ref{SWPM_simu1}(a), is due to resonant transition radiation as discussed in Refs.~\cite{chenlongRTR, chenlong2015ipac, chenlong2014aac}. Lateral THz radiation \cite{chenlong2015irmmw-thz} is also observed in the corrugated plasma channel and characterized by a coherent, narrow band spectrum as shown in Fig.~\ref{SWPM_simu1}(c). The frequencies of the first three excited THz modes based on the phase matching condition predicted by the simplified model of Sec.~\ref{dispersion} are 13.18 THz, 15.3 THz and 16.5 THz, respectively. In this case, the simulation results in Fig.~\ref{SWPM_simu1}(b) show that the THz radiation leaves the channel, creating an intensity pattern with maxima at 9 separate locations along the longitudinal distance where the fundamental mode emission dominates. 

\subsection{Dependence on Plasma Density}\label{dependence_density}
\begin{figure}
\begin{center}
\includegraphics[width=8.0cm]{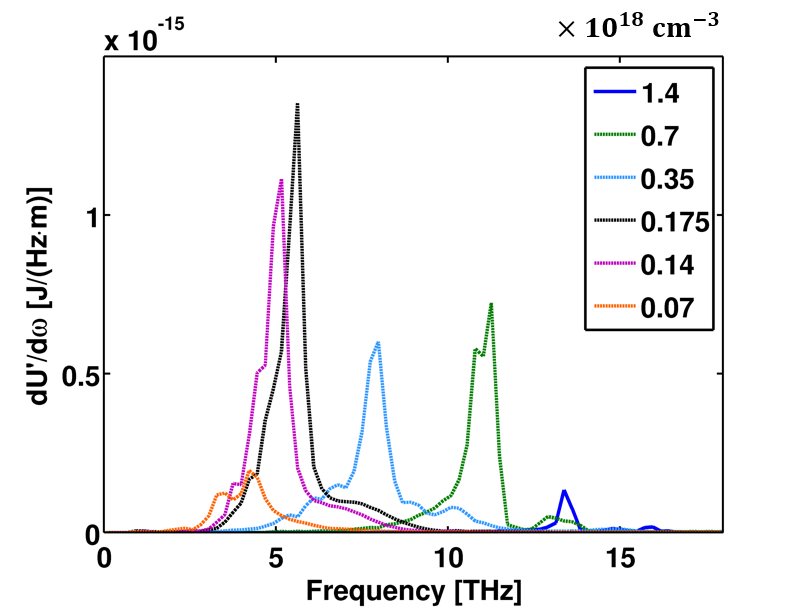}
\end{center}
\caption{(Color Online) Radiated THz spectral density $dU'/d\omega$ for different plasma densities showing dependence of central THz frequency on plasma density.} \label{SWPM_density} 
\end{figure}

We now explore the dependence of THz generation on plasma density in the case of the off-axis peak channel of Fig.~\ref{density_profile3}a. Figure~\ref{SWPM_density} displays the radiated spectral density $dU'/d\omega$, for several average, on-axis plasma densities. The central frequency of the emission peak increases with increasing density, in accord with the dependence of the channel mode frequencies on density. For all the plasma densities considered in Fig.~\ref{SWPM_density}, the fundamental mode provides the dominant contribution to the radiation energy. As shown in Fig.~\ref{SWPM_density}, the radiation generated for this channel peaks around a plasma density of $1.75\times10^{17}~cm^{-3}$. As discussed in Ref.~\cite{antonsen2007excitation}, the energy extracted from the drive pulse is converted into both electromagnetic radiation (EM) and plasma waves (PW). For example, the simulation predicts that the THz energy radiated in a 1.5 cm channel for the $1.4\times10^{17}~cm^{-3}$ density is 0.048 mJ, while the energy extracted from the laser pulse over the same distance is 4.1 mJ ($\sim$1.17\% of depletion). To efficiently deplete the laser pulse energy within a shorter channel, a higher plasma density is preferred. Therefore, to efficiently generate THz radiation, there must be a trade-off between increasing density to increase plasma current and decreasing density to increase lateral output coupling as shown in Fig.~\ref{SWPM_density}. 

\subsection{Dependence on Laser Intensity}\label{dependence_intensity}
\begin{figure}
\begin{center}
\subfigure{
 \includegraphics[width=8.0cm]{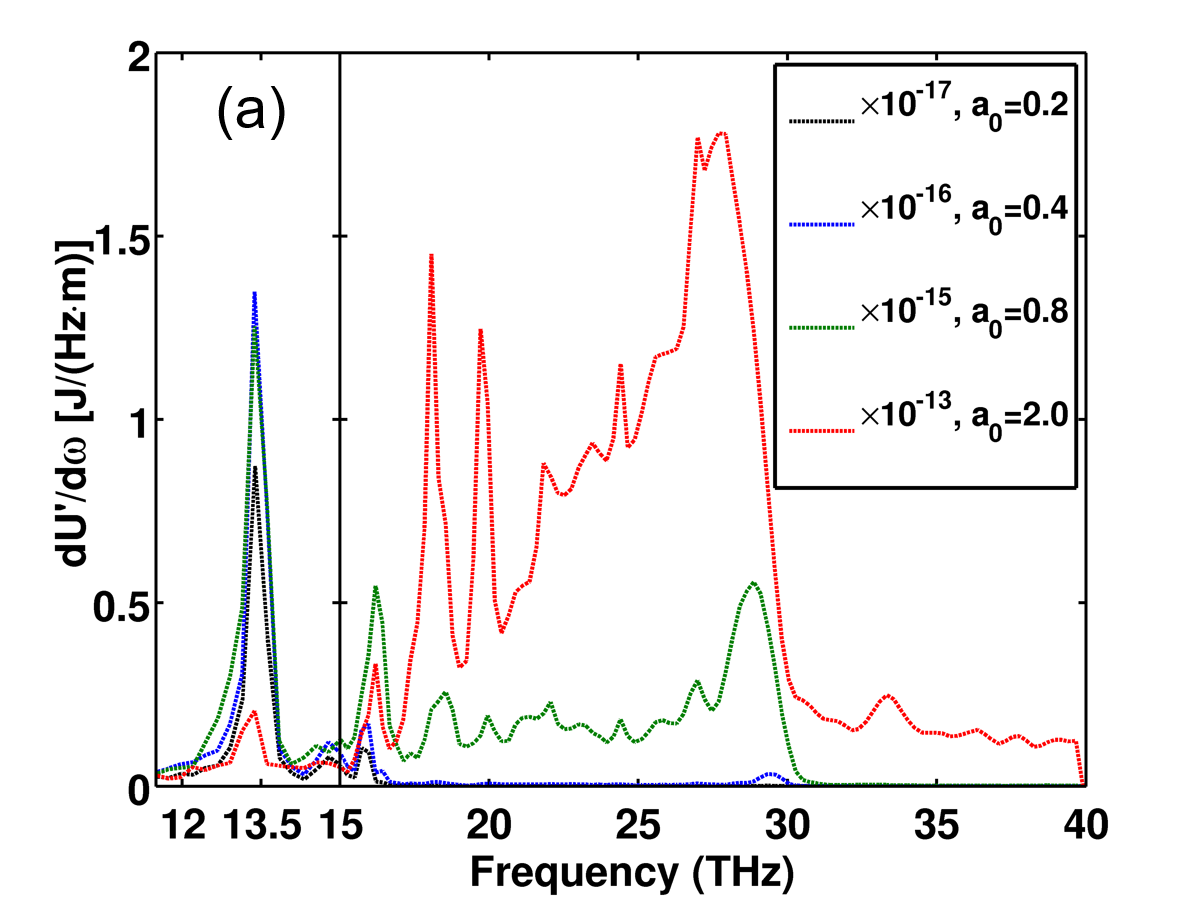}\label{spectrum_intensity}
}	
\subfigure{
 \includegraphics[width=8.0cm]{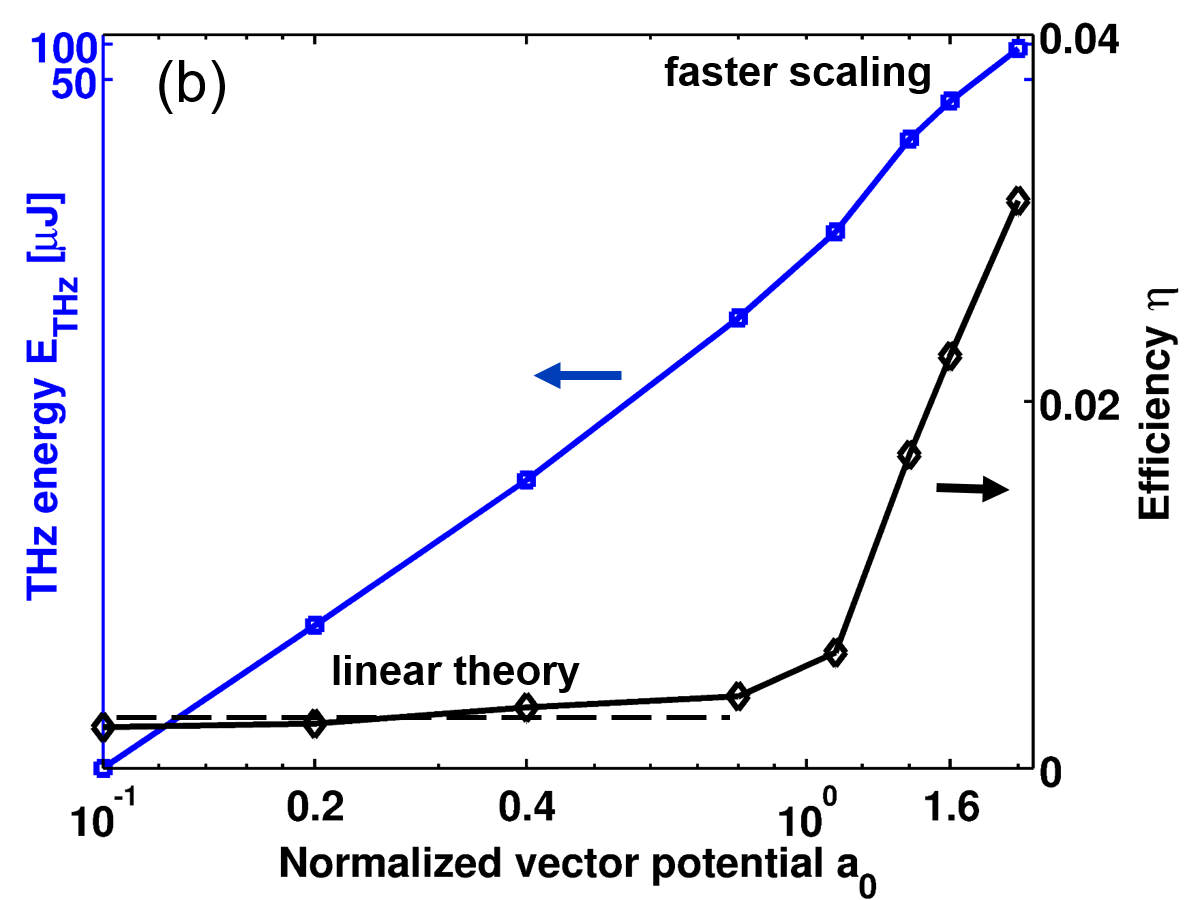}\label{efficiency_scaling_a0}
}  	
\end{center}
\caption{(Color online)  \subref{spectrum_intensity} Laterally radiated THz spectrum for different laser intensities $a_0=0.2$ (pulse energy 16.7 mJ, black), $a_0=0.4$ (pulse energy 66 mJ, blue), $a_0=0.8$ (pulse energy 0.267 J, green) and $a_0=2.0$ (pulse energy 1.66 J, red). All other parameters are given in Table \ref{table_parameters}. Note the scale factor changes as laser intensity increases. \subref{efficiency_scaling_a0} Radiated THz energy $E_{THz}$ and efficiency $\eta$ for different laser intensities. For $a_0<1$, $E_{THz}$ scales quadratically with laser intensity and $\eta$ is independent of $a_0$ as expected. However, for $a_0>1$, scaling is enhanced since higher order channel modes are excited due to nonlinear currents.}
\label{spectrum_plot_intensity} 
\end{figure}

The spatial and spectral dependences of the laterally radiated THz are illustrated for different laser intensities in Fig.~\ref{spectrum_plot_intensity}. Ponderomotively driven radiation is expected to scale with the laser amplitude as $U_L^2\sim a_0^4$ (quadratic in ponderomotive potential) for $a_0<<1$. One can see that the THz energy follows this scaling for small $a_0$, but is enhanced above this scaling for $a_0>1$. Because the relativistic ponderomotive potential scales as $a_0^2/\gamma$, the observed enhancement is even stronger than might be expected. The enhancement phenomenon is accompanied by a broadening of the THz spectrum to the point that individual modes can no longer be identified. For large $a_0$, we find that the higher frequency channel modes are excited by nonlinear currents. This, along with the broadening of the spectrum, provides the enhancement over the linear scaling. For this short channel (10 periods), the modification of the temporal profile of the laser pulse is small. Thus, the broad spectrum can be attributed to nonlinearity in the excited plasma response.

Our goal is to optimize the conversion efficiency of optical laser pulse energy to THz. Channel lengths in our simulations are limited by computation time, so we are not able to simulate every channel for a distance long enough to substantially deplete the laser pulse. We thus first consider short channels (10 periods) and define an efficiency that is the fraction of the depleted laser energy ($|\Delta E_{Laser}|$) transferred to THz, $\eta=E_{THz}/|\Delta E_{Laser}|$. By maximizing this efficiency, less power is expended driving the plasma oscillations, thus freeing it to drive THz over longer distances. We note that this efficiency does not depend on the laser intensity for $a_0<1$, based on the linear theory for which both $E_{THz}$ and $\Delta E_{Laser}$ scale as $a_0^4$. However, for large $a_0$, higher frequency THz modes are excited by nonlinear currents, enhancing the efficiency scaling. We've achieved an energy conversion efficiency of approximately 3\% for the THz generation as displayed in Fig.~\ref{efficiency_scaling_a0} and this can be further optimized by varying the corrugated plasma density profiles. For example, the conversion efficiency for the weakly relativistic case can be efficiently enhanced by finding an optimum plasma density as discussed in regards to Fig.~\ref{SWPM_density}. Further, as the laser pulse propagates through the channel, its energy depletes. The accompanying spectral red-shifting results in compression and increases the pulse amplitude \cite{zhu2013pulsed, zhu_spectral_modification2012}, which contributes to the enhancement of conversion efficiency. This phenomena will be discussed in detail in Sec.~\ref{scaling_channel_length}.

\subsection{Dependence on Laser Pulse Duration}\label{dependence_duration}
\begin{figure}
\begin{center}
 \includegraphics[width=8.0cm]{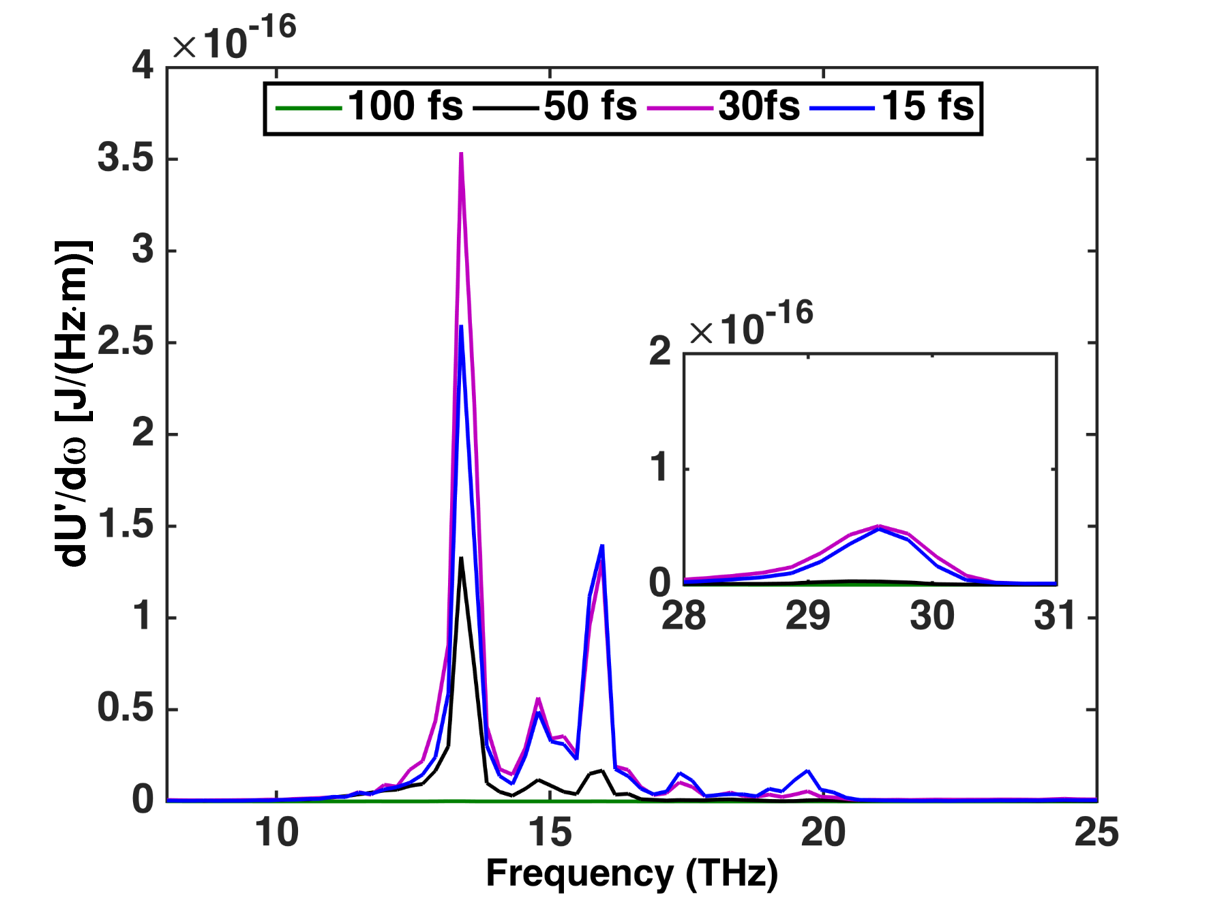}
\end{center}
\caption{(Color online)  Radiated THz spectral density $dU'/d\omega$ for different laser pulse durations, 100 fs, 50 fs, 30 fs and 15 fs, respectively.}\label{spectrum_comparison_duration}
\end{figure}

The frequencies of the linear channel modes are determined by the plasma profile. Thus, the pulse duration does not affect the THz mode frequency, but it does determine the amplitude of the driving current at each frequency. THz emission can be expected for frequencies given by the intersections of the channel dispersion curves and the laser pulse ``light line". An additional requirement for the generation of THz is that the temporal spectrum of the laser pulse envelope include the mode frequency. If the laser pulse has a Gaussian temporal profile $\exp[-(t-z/c)^2/\tau_p^2]$ with $2\sqrt{ln(2)}\tau_p$ as the pulse duration (FWHM), the amplitude of the ponderomotive driver (for fixed $a_0$) at a mode frequency $\omega$ is given by $\tau_p\exp(-\omega^2\tau_p^2/4)$, and thus minimal radiation is expected for frequencies $\omega \tau_p>>1$. Therefore, the value of $\tau_p$ can be adjusted to excite a specific range of channel mode frequencies. For excitation of the fundamental mode of 13.18 THz, the desired value is $\tau_p=16.9$ fs corresponding to a FWHM pulse duration of 28.4 fs. The simulation results further verify our estimation. Shown in Fig.~\ref{spectrum_comparison_duration} is a comparison of the radiated spectral density $dU'/d\omega$ for pulse durations of 100 fs, 30 fs, and 15 fs, respectively. The initial normalized vector potential $a_0=0.4$ is kept fixed for all 4 cases. For the case of a 100 fs laser pulse, the amplitude of the ponderomotive driver for any channel mode is small, such that minimal THz generation is observed. For the fundamental mode (13.18 THz) of the same plasma channel, the desired pulse duration is ~30 fs and the simulation result shows the THz radiation maximizes at this frequency. In addition, higher order radiation is observed as the pulse duration is shortened to 15 fs as shown in Fig.~\ref{spectrum_comparison_duration}, where there is an enhancement of the channel mode near 20 THz.

\subsection{Scaling with Channel Length}\label{scaling_channel_length}
We now investigate for the first time THz generation in corrugated plasma waveguides of sufficient length to deplete the laser pulse. As the laser pulse propagates through the plasma channel, the pulse envelope is modified and and its energy depleted through conversion into both electromagnetic radiation and plasma waves \cite{antonsen2007excitation, antonsen1992self, mora1997kinetic, shadwick2009nonlinear, PhysRevE.92.023109}. For example, the energy of a laser pulse with $a_0=2.0$ and pulse duration of 15 fs propagating through the 10 period channel shown in Fig.~\ref{simulation_setup}, is depleted by 1.12\%. At the same time the central frequency red-shifts from 375 THz to 371 THz. These changes in energy and central frequency are consistent with action conservation \cite{antonsen1992self, mora1997kinetic, zhu2013pulsed, zhu_spectral_modification2012}: as the pulse depletes, the spectrum red-shifts and the normalized vector potential $a_0$ associated with the laser pulse increases. As shown in Sec.~\ref{dependence_intensity}, the pulse depletion rate increases dramatically with intensity.

\begin{figure}
\begin{center}
 \includegraphics[width=8.0cm]{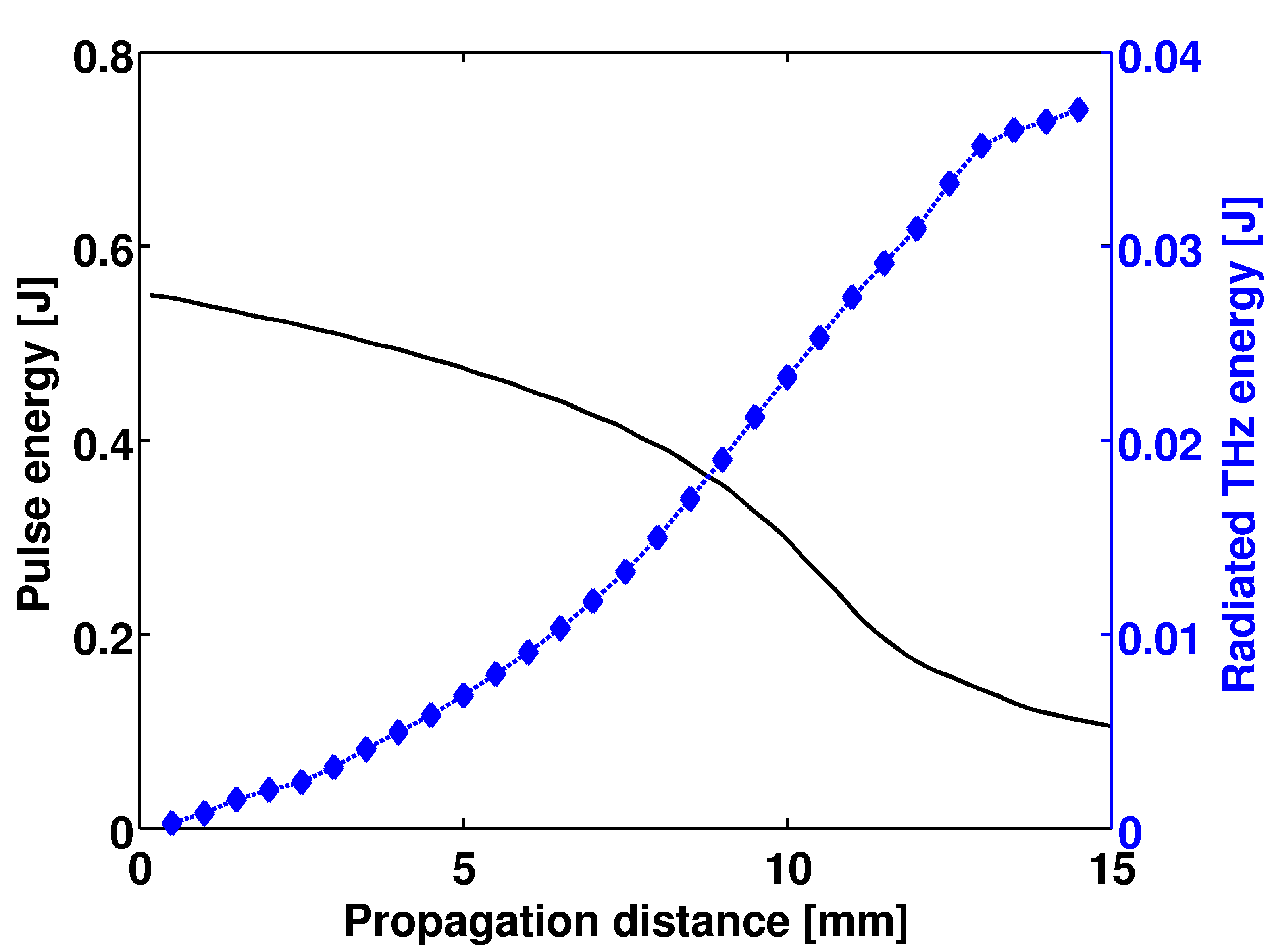}	
\end{center}
\caption{(Color online) Pulse energy depletion (black, solid) of a laser with initial $a_0=2.0$ (0.55 J) during propagation in a corrugated plasma channel. The simulation result shows that around 80\% of the energy stored in the laser pulse is depleted in a distance of 1.5 cm. Scaling of Radiated THz energy (blue, dashed) versus propagation distance shows the generated THz energy is higher than the linear scaling with distance.}\label{laser_THz_vs_channel_length_a2}
\end{figure}

We thus simulate an 800 nm laser pulse of duration 15 fs, transverse spot size of 15 $\mu$m and $a_0=2.0$ (pulse energy 0.55 J). The channel parameters are the same as in Fig.~\ref{simulation_setup}, except the channel length is extended to 1.5 cm. The simulation is conducted with a moving window of length 750 $\mu m$ to collect all the THz emission following the pulse. The 2D moving frame has a size of $205.9\times753.8$ $\mu m$ with $1024\times20480$ cells in the $x$ and $z$ directions, respectively. The energy stored in the laser pulse is displayed in Fig.~\ref{laser_THz_vs_channel_length_a2} as a function of propagation distance. Within the propagation distance of 1.5 cm, 80\% of the pulse energy is depleted. Fig.~\ref{laser_THz_vs_channel_length_a2} also displays the radiated THz energy versus propagation distance. The rate ($dU'/dz$) of THz energy generation increases with distance as the normalized vector potential $a_0$ increases during propagation due to action conservation. As a result, after propagation of 1.5 cm, more than 8\% of the total pulse energy is converted into THz radiation. 

\begin{figure}
\begin{center}
 \includegraphics[width=8.0cm]{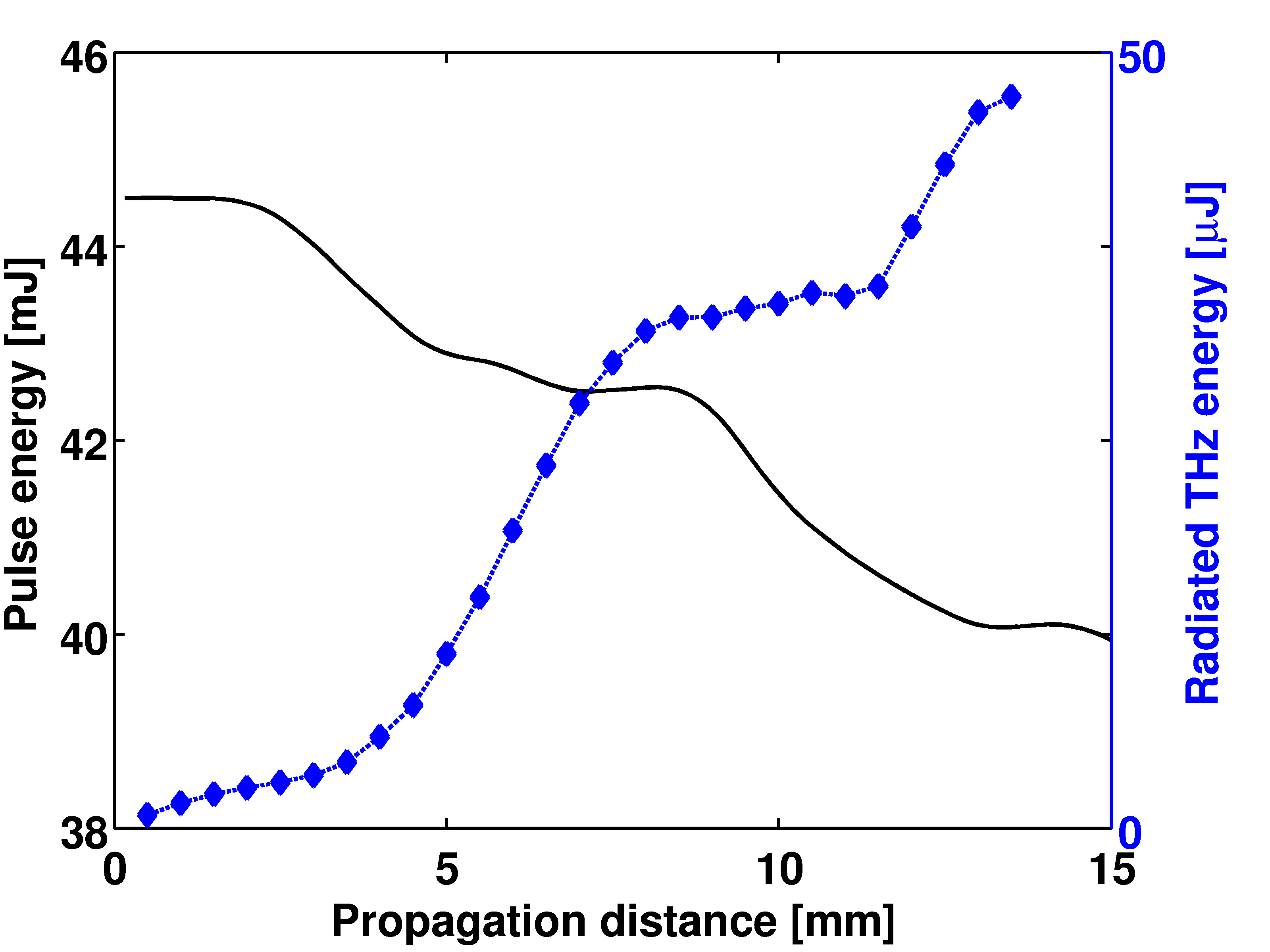}
\end{center}
\caption{(Color online) Pulse energy depletion (black, solid) of a laser with initial $a_0=0.4$ (44.5 mJ) during propagation in a corrugated plasma channel. Simulation result shows that only 10\% of the energy stored in the laser pulse is depleted in a distance of 1.5 cm. Scaling of Radiated THz energy (blue, dashed) versus plasma channel length indicates that only 1.1\% of the depleted laser energy is concerted into THz radiation in this case. }\label{laser_THz_vs_channel_length_a04}
\end{figure}

For comparison, THz generation using a relatively low intensity laser pulse with pulse duration of 30 fs, transverse spot size of 15 $\mu$m and $a_0=0.4$ (pulse energy 44.5 mJ) is also simulated.  In this case, we found the optimum plasma density for short propagation distance (10 periods) to be $1.4\times10^{17} $ $cm^{-3}$ as shown in Sec.~\ref{dependence_density}. The lower plasma density and laser intensity lead to a much longer pulse depletion length than for that of the previous case displayed in Fig.~\ref{laser_THz_vs_channel_length_a2}. Consequently, due to computational restrictions, we were not able to simulate a channel long enough to substantially deplete the pulse energy. Instead, we simulate pulse propagation through a corrugated plasma channel for a distance of 1.5 cm. The simulation results displayed in Fig.~\ref{laser_THz_vs_channel_length_a04} indicate that only 10\% of the energy stored in the laser pulse is depleted within that distance. The variation with $z$ of the depletion rate is due to the mismatch between the transverse pulse width and the matched width for the guiding channel. This leads to variations in spot size and intensity, with the depletion rate depending strongly on intensity. Shown in Fig.~\ref{laser_THz_vs_channel_length_a04}  is the radiated THz energy versus plasma channel length. About 50 $\mu$J of THz energy is generated within a 1.5 cm interaction distance. As a result, it can be concluded that the conversion efficiency for $a_0=0.4$ is much lower than that of a higher intensity pulse. However, a much longer plasma channel is needed to deplete the pulse energy for $a_0=0.4$.

\subsection{On axis peak denisty channel}\label{others}
\begin{figure}
\begin{center}
 \includegraphics[width=8.0cm]{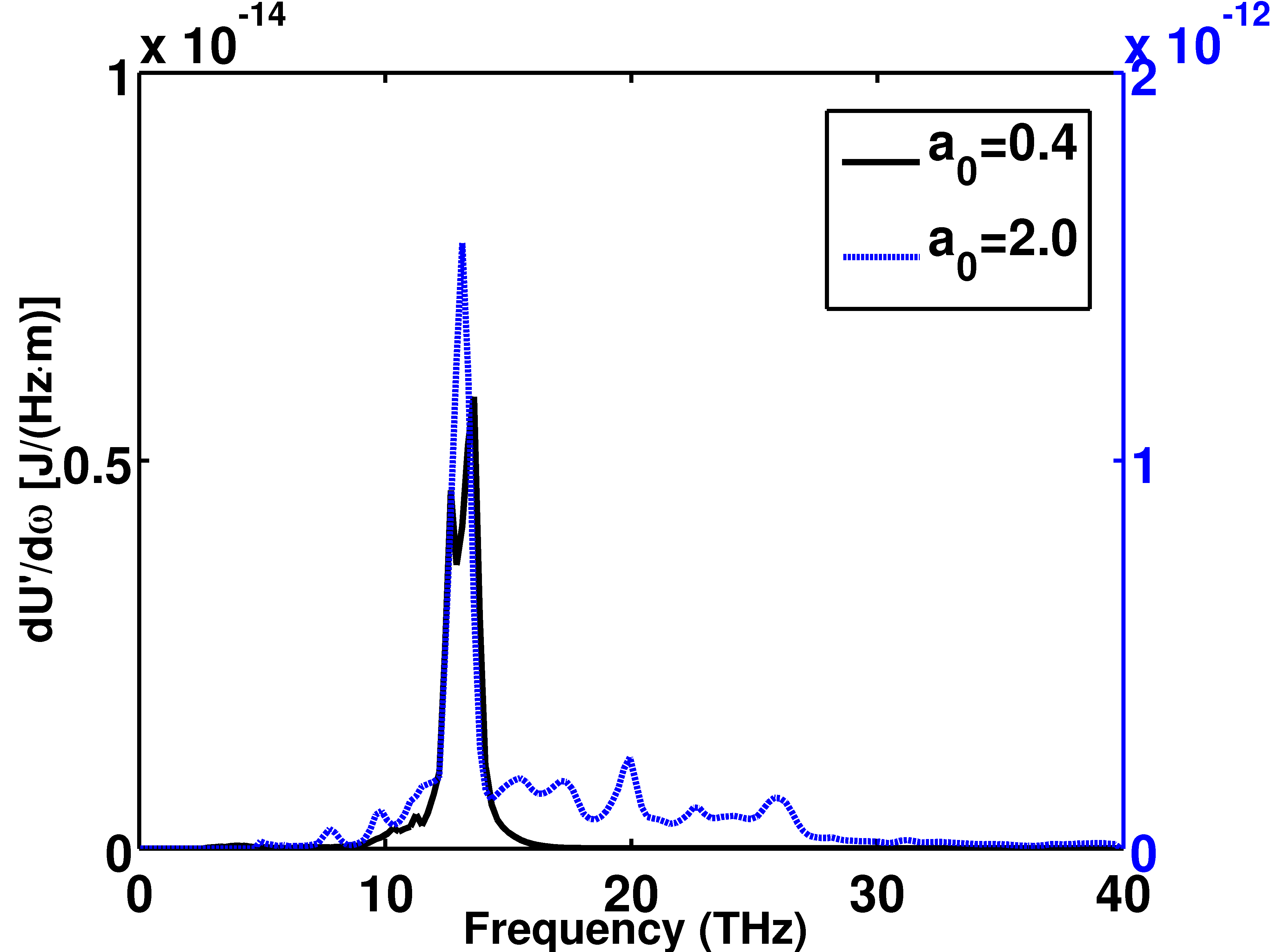}
\end{center}
\caption{(Color online) Radiated THz spectrum $dU'/dz$ for two different laser intensities $a_0=0.4$ (black, solid) and $a_0=2.0$ (blue, dashed). The plasma density profile used in the simulations is shown in Fig.~\ref{density_profile3}b and has a local on-axis peak of the channel. Simulation result shows that a narrow band THz spectrum is excited and higher THz energy is preferentially generated in this type of channel.}\label{spectrum_experiment_channel_comparison}
\end{figure}

In this section, we consider the corrugated plasma channel of the type shown in Fig.~\ref{density_profile3}b. The channel length is 10 periods with modulation wavelength of 50 $\mu$m. The other channel parameters are $\delta=0.7$, $\overline{n}_1=1.3$, $\delta_1=0.1$ (see Eq.~\ref{density_profile}). In order to match the density profile shown in Fig.~\ref{exp_setup}b, both cut-off radius and channel radius are also modulated according to $r_c[\mu m]=22.5-7.5\cos{(k_mz)}$ and $r_0[\mu m]=r_c+15$. Simulation results for laser pulses, with normalized vector potential $a_0=0.4$ and $a_0=2.0$, are shown in Fig.~\ref{spectrum_experiment_channel_comparison}. Both laser pulses have the same transverse spot size of 15 $\mu$m and pulse duration (FWHM) of 50 fs with a central wavenumber of 800 nm. The case with $a_0=0.4$ shows narrow band THz radiation is excited around the fundamental frequency of 13.6 THz. As displayed in Table~\ref{table2}, the amount of generated THz energy and conversion efficiency for this channel are significantly higher than the case shown in Fig.~\ref{SWPM_simu1} for the same laser pulse. This could be explained by the excitation of a higher electron current due to the relatively greater density inhomogenity experienced by the driver. In addition, the axially averaged density profile has a lower radial barrier that allows the generated THz waves to escape the channel. For a higher intensity laser pulse with $a_0=2.0$, the generated THz shown in Fig.~\ref{spectrum_experiment_channel_comparison} is characterized by a different spectrum compared with Fig.~\ref{spectrum_intensity}. Although the amount of THz energy is still enhanced relative to $a_0=0.4$, the spectrum is confined in a relatively narrow band near the fundamental frequency while in the case of Fig.~\ref{spectrum_intensity}, higher order THz modes are significantly generated and consequently modify the spectrum.

\section{Conclusions and Discussion}\label{conclusion}
We have investigated THz generation in corrugated plasma channels accounting for nonlinear excitation of plasma waves and laser pulse depletion. Theoretical analysis and Full format PIC simulations were conducted. A range of laser pulse parameters and plasma channel structures were considered with the goal of maximizing the conversion efficiency of optical pulse energy to THz energy.

Table~\ref{table2} displays the simulation results for different pulse and plasma parameters for the two types of corrugated channel displayed in Fig.~\ref{density_profile3}. Most of the simulations were conducted using a 10-period channel to investigate the conversion efficiency of the depleted optical pulse energy to generated THz. For these simulations, the pulse energy only depleted a small percentage. Three examples of longer channels were conducted to examine the consequences of significant pulse depletion and the results are included in Table~\ref{table2}. Our general conclusions are as follows. Generally THz generation increases with laser amplitude $a_0$. For fixed $a_0$, THz generation at frequency $\omega$ maximizes where $\omega \tau_p \sim 1$, where $\tau_p$ is the pulse duration. This applies for both channel types. For a short channel with only 10 periods (channel length 0.5 mm), simulation results indicate more THz energy is generated for the on-axis peak density channel type shown in Fig.~\ref{density_profile3}b. Efficient THz generation involves a trade-off between increasing density to increase the plasma current and decreasing density to increase the lateral output coupling. However, since the excited THz mode depends on the channel structure according to the dispersion relation discussed in Sec.~\ref{dispersion}, lower plasma density also modifies the generated THz spectrum. In addition, to efficiently deplete the laser pulse energy within a shorter channel, a higher plasma density is preferred. 

As an example, we choose a laser pulse with $a_0=2.0$ and pulse duration of 30 fs. The channel type shown in Fig.~\ref{density_profile3}b is used with an averaged on axis electron density of $1.4\times10^{18}$ $cm^{-3}$ and the channel length is extended to 1.5 $cm$. The simulation results as displayed in Table~\ref{table2} show after the laser pulse propagates through the channel for 1.5 $cm$, around 48\% of the laser pulse energy is depleted and 16 mJ of this energy is converted to THz energy with a narrow spectrum (Fig.~\ref{spectrum_experiment_channel_comparison}). The conversion efficiency is around 3\% and less than the case shown in Sec.~\ref{scaling_channel_length}. This is probably due to the fact that the channel displayed in Fig.~\ref{density_profile3}b no longer remains a transverse parabolic structure capable of guiding. In fact, laser energy leaks laterally and the wave action is not conserved as the laser pulse propagates through the channel. Therefore, the normalized vector potential $a_0$ decreases with the propagation distance and less THz energy is generated.

\begin{table}
\begin{center}
\caption{Comparison of generated THz energy for different laser intensities and density profiles.}\label{table2}
\begin{tabular}{c|c*{6}{c}}
\hline \hline
\multirow{2}{*}{Channel Type} &$a_0$& pulse duration &on-axis density&channel length& energy depletion & THz energy &efficiency  \\
&  & [fs] & [$\times 10^{18}$ $cm^{-3}$] & [mm] & into the channel& [mJ] & $\eta$  \\
\hline
\multirow{4}{*}{Fig.~\ref{density_profile3}a} & 0.4 & 30 &0.14 &15 &10.11\% & 0.048 & 1.11\%\\
& 0.4 & 50 &1.4 & 0.5&  0.0175\% & 3.36e-5 & 0.28\%\\
& 2.0 & 15 & 1.4& 15 & 81.8\% & 38 & 8.44\%\\
& 2.0 & 50 & 1.4& 0.5& 0.39\% & 0.2 & 3.25\%\\
\hline
\multirow{5}{*}{Fig.~\ref{density_profile3}b} & 0.4 & 30 &1.4 &0.5 & 0.033\% & 0.0016 & 12.03\%\\
& 0.4 & 50 &1.4 & 0.5 & 0.024\% & 0.0013 & 8.17\%\\
& 0.4 & 50 &0.14 & 0.5 & 0.006\% & 4.68e-4 & 11.7\%\\
& 2.0 & 30 & 1.4 & 15 &  48.2\% & 16 & 3\%\\
& 2.0 & 50 & 1.4 & 0.5 &  0.4\% & 0.6493 & 9.84\%\\
\hline \hline
\end{tabular}
\end{center}
\end{table}

In order for the present mechanism to be a useful high power source of THz radiation, the spectrum should be tunable. Since the excited THz consists of a superposition of the channel modes of the corrugated plasma structure, all the parameters used in Eq.~(\ref{density_profile}) can be tuned to modify the THz spectrum. More specifically, varying the averaged on axis density $N_{0}$ is the most straightforward way to tune the frequency in the experiment. For example, Figure.~\ref{SWPM_density} shows that varying the average density $N_{0}$ from $1.75\times10^{17}$ $cm^{-3}$ to  $1.4\times10^{18}$ $cm^{-3}$ results in a shift in the central frequency from 5 THz to 14 THz. For the channel type of Fig.~\ref{density_profile3}a, the THz spectrum changes dramatically with laser intensity going from a narrow spectrum at $a_0=0.4$ to a broad spectrum at $a_0=2.0$ with enhanced energy as well. For the channel type of Fig.~\ref{density_profile3}b, the spectrum remains strongly peaked over the same range of laser amplitudes.  Since the generated THz waves are emitted laterally, one needs an optical system to collect the radiation. This might be realized with a conical mirror to focus the THz radiation to one direction for practical uses. Overall, the mechanism, using realistic corrugated plasma structures, presented in this paper provides a potential high power source of THz with a tunable spectrum and a conversion efficiency of over 8\%.


\acknowledgments{The authors would like to acknowledge Dr. Daniel Gordon for the use of TurboWAVE and thank Luke Johnson and Thomas Rensink for fruitful discussions. Part of the simulations was performed on NERSC Edison and Deepthought2 clusters at UMD. This work was supported by the Office of Naval Research and the US Department of Energy. }

\appendix
\section{Calculating frequencies of the radial eigenmodes}\label{appendix1}

In this appendix, we will discuss the numerical method to calculate the exact frequencies of the axially averaged excited modes. Equation~(\ref{wave_equation}) can be further written as
\begin{equation}
\label{wave_equation_02}
\left(\frac{1}{r}\frac{\partial}{\partial r}r\frac{\partial}{\partial r}-\frac{1}{r^2}\right)E_r
+\left(k_c^2-k_p^2(r)\right)E_r=0\; ,
\end{equation}
where the cut off wavenumber $k_c$ of EM modes and plasma wavenumber $k_p$, are defined as, $k_c^2=\omega^2/c^2-k_0^2$ and $k_p^2(r)=\omega_p^2(r)/c^2$, respectively. Since we consider radially polarized TM modes, $E_r$ must vanish on axis, i.e., $E_r(0)=0$. We know that outside the channel, $N(r>r_0)=0$, thus Eq.~(\ref{wave_equation_02}) yields to Bessel's differential equation. The far field $E_r$ outside the channel must match the properties of an outgoing wave, which has the form of the first kind Hankel function $H_1^{(1)}(k_c r)$ and asymptotically behaves as $E_r \sim \frac{1}{\sqrt{r}}\exp{(ik_c r)}$. The boundary condition allows us to know the ratio of $E_r$ to its derivative outside the channel. As a result, we can numerically integrate Eq.~(\ref{wave_equation_02}) using the shooting method to determine the $k_c$ values that satisfy the on axis boundary condition $E_r(0)=0$.

To calculate the radial eigenmodes numerically for an arbitrary but given transverse density profile, one can numerically evaluate $E_r$ by Eq.~(\ref{wave_equation_02}) using the shooting method for a set of $k_\perp$ and determine what $k_\perp$ satisfies the on axis boundary condition $E_r(0)=0$. For mathematical simplicity we set $\Phi=rE_r$, $\beta(r)=k_p^2(r)$ and Eq.~(\ref{wave_equation_02}) yields to
\begin{equation}
\label{wave_equation_03}
\frac{\partial^2\Phi}{\partial r^2}-\frac{1}{r}\frac{\partial \Phi}{\partial r}+\left(k_c^2-\beta(r)\right)\Phi=0\; .
\end{equation}

One can also find as $r\to 0$, $\Phi \sim a\cdot r^2+b$; the on axis boundary condition is satisfied only if $b\to0$. The finite difference (FD) shooting method can be implemented by 
\begin{equation}
\label{FD_shooting}
\Phi_{j-1}=-\frac{\Phi_{j+1}(2j-1)+2j\big[h^2(k_c^2-\beta_j)-2\big]\Phi_j}{1+2j}\; ,
\end{equation}
where $h$ is the step size.

\begin{figure}
\begin{center}
\includegraphics[width=10.0cm]{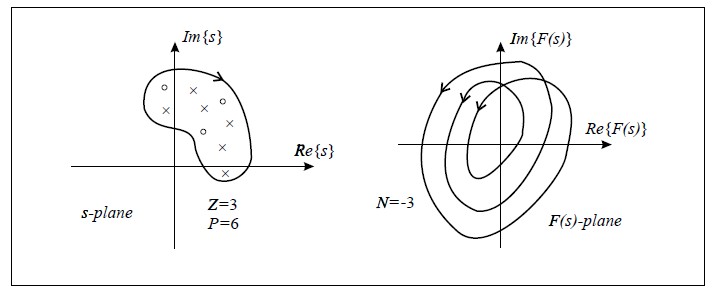}
\end{center}
\caption{Nyquist Theory: Cauchy's principle.}\label{nyquist}
\end{figure}

To find out the number of different radial modes, i.e. $k_c$, which can be supported by a channel with finite transverse size, one can scan the parameter $k_c$ in Eq.~(\ref{FD_shooting}) and apply the Nyquist Theory illustrated in Fig.~\ref{nyquist}. $F(s)$ is an analytic function defined in a closed region of the complex $s$-plane shown on the left. As $s$ travels a clockwise path in the $s$-plane, $F(s)$ encircles the origin on the complex $F(s)$-plane $N$ times,
\begin{equation}
N=Z-P\; ,
\end{equation}

where $Z$ and $P$ denote the number of zeros and poles of the function $F(s)$ in the closed region, respectively. For our shooting method, as shown in Eq.~(\ref{FD_shooting}), $\Phi(0)$ has no poles and the result yields $N=Z$.

\begin{figure}
\begin{center}
\includegraphics[width=10.0cm]{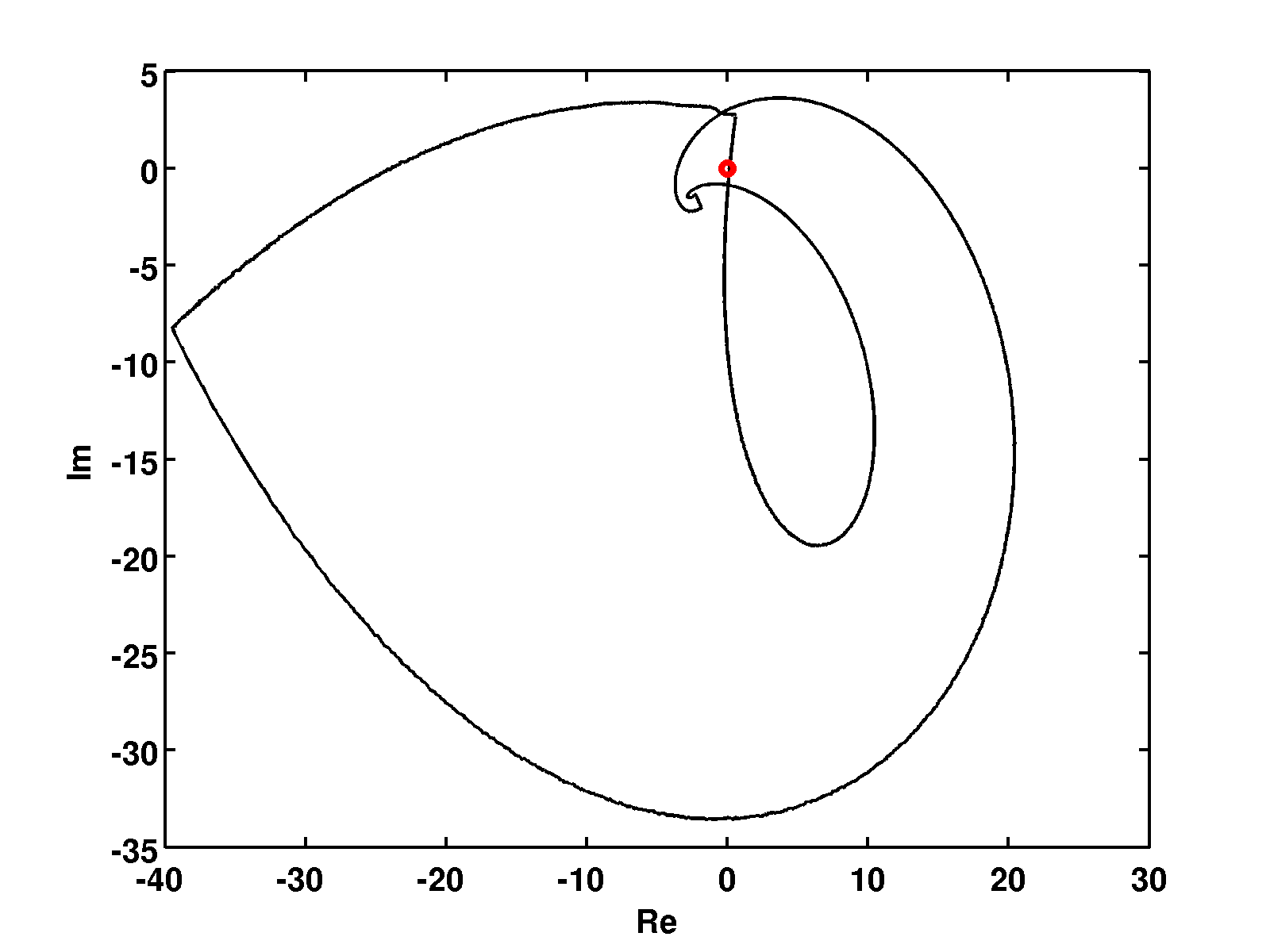}
\end{center}
\caption{(Color Online) $\Phi(0)$ encircles origin (red dot) twice as $k_c$ scans from 0.301 $\mu m$ to 0.402 $\mu m$. The density profile in this case is: $N_{0}=1.4\times10^{18} ~cm^{-3}$, $\delta_1=\delta=0.9$, $\overline{n}_1=3$, $r_c=30~\mu m$ and $r_0=40~\mu m$.}\label{nyquist_total}
\end{figure}

\begin{figure}
\begin{center}
\subfigure{
 \includegraphics[width=8.0cm]{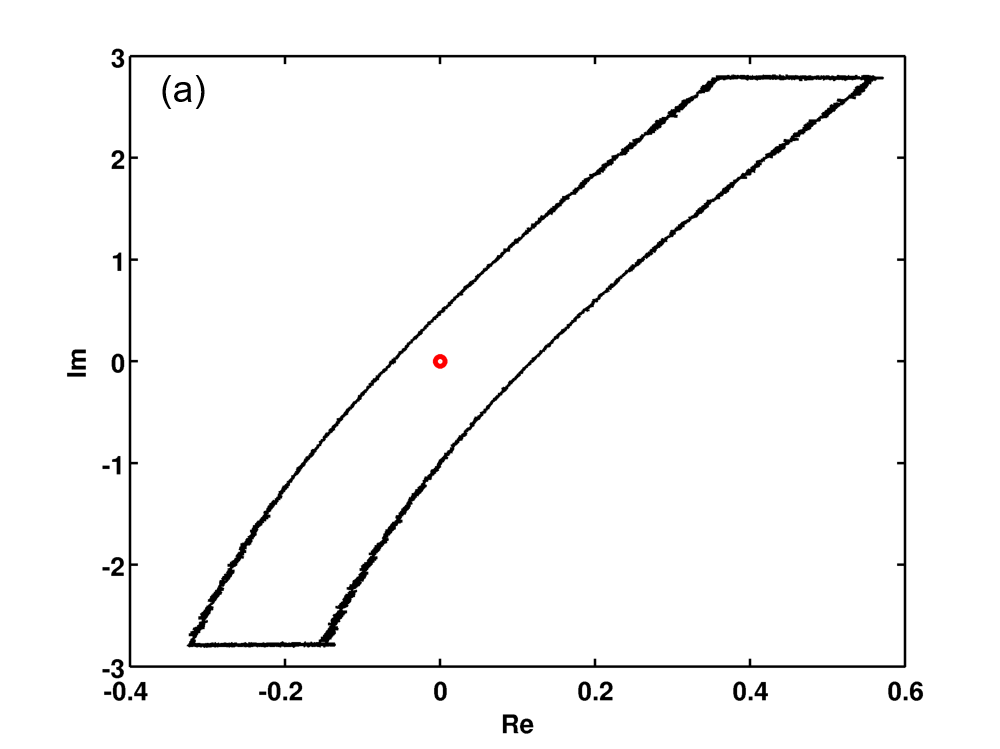}\label{1st_mode}
}
\subfigure{
 \includegraphics[width=8.0cm]{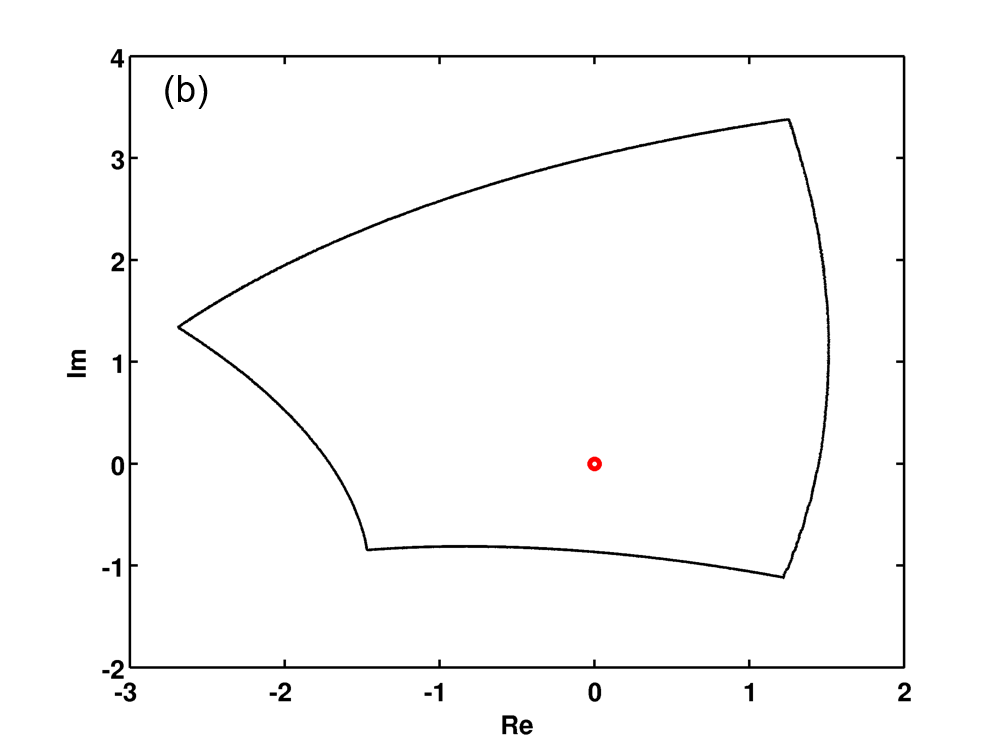}\label{2nd_mode}
}  	
\end{center}
\caption{(Color online) \subref{1st_mode} $k_c$ of the fundamental radial mode is found to be between $0.301~\mu m^{-1}$ and $0.303~\mu m^{-1}$.  \subref{2nd_mode} $k_c$ of the second order radial mode is found to be between $0.36~\mu m^{-1}$ and $0.37~\mu m^{-1}$. Applying linear interpolation can help narrow the range of $k_c$ and we find $k_c=0.30195~\mu m^{-1}$ for the first radial mode and $k_c=0.3638~\mu m^{-1}$ for the second radial mode.}
\end{figure}

This method predicts the frequency of radiation for any given transverse density profile. For example, Fig.~\ref{nyquist_total} shows for a density profile shown in Fig.~\ref{simulation_setup}, $\Phi(0)$ encircles the origin twice as $k_c$ scans from 0.301 $\mu m^{-1}$ to 0.402 $\mu m^{-1}$, which implies that the channel can support two radial eigenmodes according to Cauchy's principle. Further one can apply linear interpolation to narrow the range of $k_c$ for each mode to find the exact value of $k_c$ for the field to satisfy the boundary condition. Figures~\ref{1st_mode} and \ref{2nd_mode} are two figures indicating the range of $k_c$ during the interpolation to find the exact value $k_c$ for first and second radial eigenmodes, respectively.

The calculation of $k_c$ for the radial eigenmodes is displayed in Table \ref{table} and matches closely with the estimate $k_c=\sqrt{\omega_{p0}^2/c^2 +8\gamma/w_{ch}^2}$ from Eq.~(\ref{channel_disp}).
\begin{table}
\begin{center}
\caption{Exact values of numerically calculated $k_c$ and estimation from Eq.~(\ref{channel_disp}) for different radial modes.}\label{table}
\begin{tabular}{ccc}
\hline \hline
Mode number& $k_c[\mu m^{-1}]$, exact &  $k_c[\mu m^{-1}]$, Eq.~(\ref{channel_disp}) \\
\hline
1st & 0.30195 & 0.3024\\
2nd & 0.36384 & 0.3652\\
\hline \hline
\end{tabular}
\end{center}
\end{table}

\bibliography{thz_manuscript}

\end{document}